\documentclass[sigconf,authorversion, nonacm]{acmart}

\AtBeginDocument{%
  \providecommand\BibTeX{{%
    \normalfont B\kern-0.5em{\scshape i\kern-0.25em b}\kern-0.8em\TeX}}}

\setcopyright{acmcopyright}
\copyrightyear{2024}
\acmYear{2024}
\acmDOI{XXXXXXX.XXXXXXX}

\usepackage{multirow}

\title{Here Be Livestreams: Trade-offs in Creating  Temporal Maps of Reddit}
\author{Virginia Partridge}
\affiliation{
    \institution{University of Massachusetts Amherst}
    \city{Amherst}
    \state{Massachusetts}
    \country{USA}
}
\email{vcpartridge@umass.edu}

\author{Jasmine Mangat}
\affiliation{
    \institution{University of Massachusetts Amherst}
    \city{Amherst}
    \state{Massachusetts}
    \country{USA}
}
\email{jmangat@umass.edu}

\author{Rebecca Curran}
\affiliation{
    \institution{University of Massachusetts Amherst}
    \city{Amherst}
    \state{Massachusetts}
    \country{USA}
}
\email{rebecca@mediacloud.org}

\author{Ryan McGrady}
\affiliation{
    \institution{University of Massachusetts Amherst}
    \city{Amherst}
    \state{Massachusetts}
    \country{USA}
}
\email{ryan@mediacloud.org}

\author{Ethan Zuckerman}
\affiliation{
    \institution{University of Massachusetts Amherst}
    \city{Amherst}
    \state{Massachusetts}
    \country{USA}
}
\email{ethanz@umass.edu}

\begin{document}

\begin{abstract}
We present a method for mapping Reddit communities that accounts for temporal shifts, using quantitative and qualitative analyses of clustering techniques to produce high-quality, stable, and meaningful maps for researchers, journalists and casual Reddit users. Building on previous work using community embeddings, we find that only a month of Reddit comments suffices to create snapshot embeddings that maintain quality while supporting insight into changes in Reddit communities over time. Comparing different clusterings of community embeddings with quantitative measures of quality and temporal stability, we describe properties of the models and what they tell us about the underlying Reddit data. Moreover, qualitative analysis of the resulting clusters illuminate which properties of clusterings are useful for analysis of Reddit communities. Although clusterings of subreddits have been used in many earlier works, we believe this is the first study to qualitatively analyze how these clusterings are perceived by social media researchers at a Reddit-wide scale. 

 Finally, we demonstrate how the temporal snapshots might be used in exploratory study. We are able to identify particularly stable communities during 2021-2022, such as the Reddit Public Access Network, as well as emerging communities, like one focused on NFT trading. This work informed the development of a webtool for exploring Reddit now available to the public at \url{RedditMap.social}.
\end{abstract}


  

\maketitle

\section{Introduction}
The social media site Reddit consists of thousands of discrete, self-organized, topic-specific forums called subreddits. Individual users known as Redditors can post text, images, videos, or links and in turn comment and vote on other users' contributions\cite{Medvedev2018-so}. Users subscribe to subreddits covering any interests, ranging from general topics, such as \textsf{AskReddit}\footnote{Subreddits indicated with \textsf{teletype font} throughout.}, to those focused on specific hobbies, political ideologies, or communities with physical counterparts, like universities or cities. Rather than digitally connecting two people who have already met in a school or workplace, Reddit connects strangers who have common interests by way of subreddits\cite{IllustratedGuide}. In recent years, the site has garnered attention in news media and academic research focused on topics ranging from analyzing political polarization and partisanship\cite{Waller2021-yq, RedditAndStruggleToDetoxify} to stock and option trading in the subreddit \textsf{wallstreetbets}\cite{Hasso2021WhoPI, MisfitsShakingWallStreet}. 

Reddit's moderation policies also distinguish it from other platforms. Subreddit community moderators create and enforce their own rules with little guidance from the platform, although subreddits violating Reddit's company policies on spam, anti-harassment, hate-speech or illegal content are banned or "quarantined" (hidden from unsubscribed users)\cite{Chandrasekharan2017-he, singh2019,  reddit_help_quarantined_subreddits}. As a result, community norms, volunteer efforts and self-policing all influence participation and content on the site. This community-based approach to moderation, coupled with easy access to years of data thanks to the Pushshift Reddit Dataset\footnote{Note that Reddit recently removed Pushshift's access to their Data API, \url{https://www.reddit.com/r/modnews/comments/134tjpe/reddit_data_api_update_changes_to_pushshift_access/}}\cite{Baumgartner2020-vo}, has made Reddit a fertile ground of study for anyone interested in changes in discourse on social media and how communities evolve.

As the role of social media in civic life grows, the need for tools to explore and understand this space has also increased. We seek to build a temporal map of Reddit legible not only to academics, but also journalists and activists seeking to inform the public about trends in social media communities and Reddit users looking to explore new communities. Our tool contextualizes highly visible communities, targets of recent media attention like \textsf{wallstreetbets}, among many equally active and vibrant communities worthy of study, such as those dedicated to hobbies, self-help, and fandom, to name just a few. We strive to provide insights into how online communities change over time, letting researchers see when subreddits shift in response to offline events or interactions within the Reddit community. In this paper, we lay out the methodologies to create this tool, characterize trade-offs made along the way, examine statistical properties of subreddit clusterings, and validate the usability of the tool from the perspective of social media researchers. 

While building our map of Reddit, we draw inspiration from the variety of ways Earth's globe can be mapped. Taking a 3-dimensional sphere into 2-dimensions necessarily means information is distorted, and so different types of maps have different purposes and trade-offs. The familiar Mercator projection is useful for navigation, but distorts relative sizes, making land areas near the Arctic appear much larger than they actually are\cite{map_projections}. Which lines and labels are drawn also matters; lines marking boundaries on a political map serve a different purpose from lines in a topographic map. The variables are endless, but cannot all be included without overwhelming users. As we build models to map online spaces, we make similar choices, distilling millions of posts and comments into categories of communities and the relationships between them. 

Drawing on recent work with community embeddings for Reddit, we propose a method for mapping subreddits that also accounts for temporal shifts by using monthly snapshots of data. With relatively small amounts of data, we create monthly snapshot community embeddings that infer relationships between subreddits, allowing clustering as an intuitive way to explore Reddit\footnote{Code and annotated data are available at \url{https://github.com/UMassCDS/IHOP-Reddit}}, but also adaptable to other forum-based platforms. Through annotation by human experts and intrinsic evaluation metrics, we validate that these monthly community embeddings consistently produce high quality, understandable clusters, although users find kmeans++ clusterings more intuitive than hierarchical agglomerative clustering.  

Our approach also reveals areas of change and stability in Reddit communities over time in a way that supports both exploratory and confirmatory analyses. No community is static. Subreddits may be impacted by external influences, such as an explosion in new users due to media attention, platform-wide changes in Reddit's interface or policies, or internal changes in community norms like moderation rules. We apply \textit{variation of information} to see that most changes in clusterings are gradual over time, while the stability of subreddits' nearest neighbors in the embedding space reveals insights about how specific communities change from one time period to the next. Additionally, representations of Reddit during each time period are produced independently from previous time periods, retaining the ability to make comparisons across time without the overhead of reprocessing years of data. 

\section{Related Work}

Launched in 2005, Reddit hosts self-organizing and self-moderating forums, where users can create and participate in subreddits that align with their personal interests, identity and preferred rules of engagement. It is a popular source for studying diverse features of online communities such as community creation, connections and conflicts between different subreddits, and patterns of radicalization and political polarization\cite{Medvedev2018-so, Tan2018TracingCG, Kumar2018CommunityIA, Grover_Mark_2019, Waller2021-yq}. Methods of study have been equally diverse, ranging from case studies on a small set of subreddits\cite{Thach2022InvisibleMA, balci_beyond_fish_and_bicycles}, to vector space models or topic models with features derived from comment text and user metadata\cite{Martin2017community2vecVR, Zhang2017CommunityIA, Kumar2018CommunityIA}, to network backbone extraction and community detection\cite{Olson2015NavigatingTM}.

Many approaches to mapping Reddit rely on learning a notion of distributional similarity for subreddits based on user behavior, perhaps understood more simply as "birds of a feather flock together". An early large scale interest map of Reddit used the inferred link that occurs between two subreddits when a single user posts to both during a particular time period. Backbone network extraction was applied to those inferred connections between subreddits in order to visualize Reddit circa 2013 as a network graph with 59 distinct \textit{interest meta-communities} clusters, such as \textit{sports}, \textit{programming} and \textit{general interest}. Their goal was to provide users with a way to find new subreddits matching their interests via a browse-able network graph exploration tool\cite{Olson2015NavigatingTM}. Backbone network extraction excels at highlighting strengths of relationships between nodes in social networks. However, the method is sensitive to algorithmic hyperparameters and filtering settings, and results are often difficult to interpret and evaluate, making it challenging to maintain consistent, comparable maps over time\cite{Neal2014, Gursoy2021ExtractingBackbone}.

\begin{figure}[t]
\centering
\includegraphics[scale=0.5]{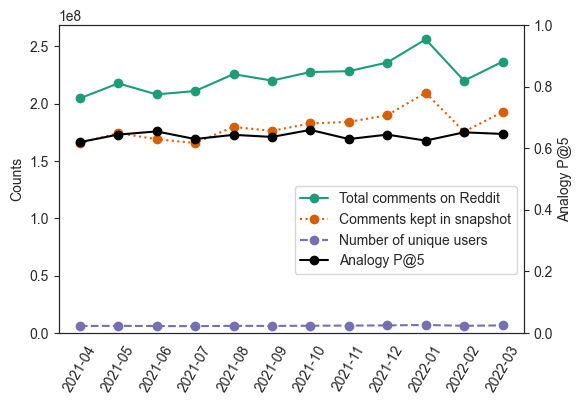}
\caption{Total number of comments, unique user contexts from each month's data snapshot, and Precision@5 performance of the best community embedding trained on that snapshot. P@5 performance each month is consistently high, averaging 0.64 (std. dev. 0.012) and never drops below 0.61.}
\label{fig:snapshotstats}
\end{figure}

A Redditor's comments across different forums can also be used to learn dense vector representations of subreddits, the approach we will follow here. This method, first introduced as \textit{community2vec}\cite{Martin2017community2vecVR}, is based on word embeddings and, like its word2vec and GloVe predecessors with natural language, is capable of solving meaningful analogies and generating useful similarity relationships in the subreddit vector space\cite{Martin2017community2vecVR, Waller2021-yq, Mikolov2013EfficientEO, pennington2014glove}. For example, GloVe-based embeddings of subreddits were shown to successfully retrieve the corresponding sports team when presented with a league and geographic location, \textsf{warriors = sanfrancisco + nba}\cite{Martin2017community2vecVR}. This insight led to employing sets of analogies to tune subreddit embeddings, which researchers used to examine which Redditors and communities focus on broad, general interests compared to those that specialize in particular topics and which communities exhibit age, gender and political biases\cite{Waller2019GeneralistsAS, Waller2021-yq}. 

The resulting embedding space can be explored by looking at the neighbors of \textit{seed} subreddits or clusters of subreddits, suggesting an iterative discovery process where related communities can be found for comparisons and case studies. Balci et al.\cite{balci_beyond_fish_and_bicycles} used subreddit embeddings from 2019 Reddit data in this way to determine which subreddits engaged with discourse on feminism and gender, ultimately constructing a taxonomy of online women's ideological spaces. However, this type of study requires embedding models built with data from the time period of interest, and we found that previous work on clustering Reddit relied on long time periods of data, using at least one year's worth of Reddit data to build a single model and often more than five years. In such models, changes over time are obscured, as they are "averaged out" to a single vector weight in the final model. Including more recent data requires training a new model over the entire time frame, in which case studying temporal trends demands different methods, such as snapshot language models\cite{Zhang2017CommunityIA} or genealogy graphs\cite{Tan2018TracingCG}. Not everyone has access to the computing infrastructure to train their own embeddings and clusters. We aim to make subreddit embeddings more accessible to the research community, offering a way to compare temporal snapshots to see subreddits' evolution, temporal trends, and relationships between Reddit communities without training from scratch or sacrificing the quality and utility of community embeddings established by previous authors.

\section{Datasets}

\subsection{Reddit Comments}
We use the comments portion of the Pushshift Reddit Dataset from April 2021 through March 2022\cite{Baumgartner2020-vo} to create our monthly snapshots. Within each $t$-th month, we determine the top 10,000 subreddits by number of comments, removing any other subreddits and user profile pages. Even the least active subreddits included had at least 1750 comments in any single month of our dataset. We intentionally selected the top 10K subreddits both because it is the threshold used in earlier work\cite{Waller2021-yq} and because Redditors interacting in popular subreddits tacitly understand they may get significant public attention. We acknowledge the potential for our tool to draw unwanted attention and create an observer effect, where attention might change the way people use Reddit. This attention may be quite harmful to users if it results in them abandoning subreddit communities or long-held pseudonyms to avoid embarrassment or scrutiny. In many spaces on the internet, such harmful attention disproportionately affects marginalized communities. The fact that community embeddings do not expose usernames, comment text or any other information that could be used to identify a user make them appealing from an ethical and privacy perspective\cite{Reagle2022DisguisingRS}. Including only the most actively commented on subreddits also mitigates these risks of harm.

After selecting only the top 10K subreddits, we drop comments from deleted users, users that only commented once on Reddit during the month and comments that were themselves deleted or removed. As a final step, we remove users above some portion of the most active remaining users during that time period, again by number of comments. Although initially intended as a heuristic to remove bots and spam, we found that the percentile could be tuned to increase accuracy on the subreddit analogy tasks. In initial experiments, removing users above the 95th percentile of most active monthly users performed well and is the strategy adopted for this work. This translates an average maximum of 74 comments by a single unique user and average 28 comments per user across the months in our dataset. As a beneficial side effect, removing the most prolific users also speeds up training embeddings by excluding the longest contexts from the data. 

Finally, the names of subreddits each user commented on during the time period are collected into that user's context. All user contexts make up the data contributing to our snapshot of Reddit at month $t$.  Count statistics of comments and unique users for each $t$-th snapshot can be seen in figure \ref{fig:snapshotstats}. Overall, 6,950 subreddits appear in every month's snapshot, while 15,292 appear in at least one month.  

\begin{figure}[t]
    \centering
    \includegraphics[scale=0.6]{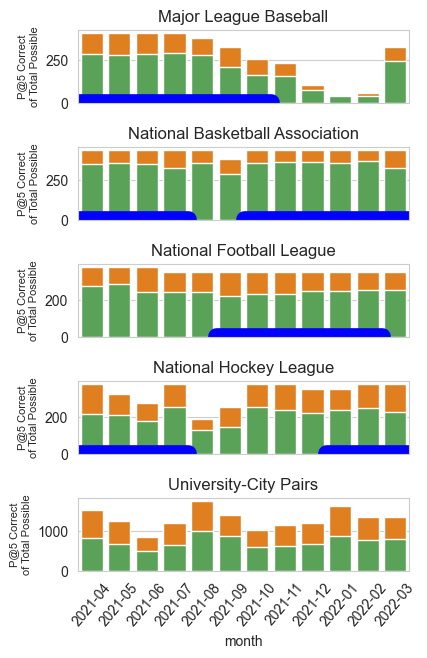}
    \captionof{figure}{Proportion of analogies solved according to Precision@K out of the total solvable in top 10K most commented subreddits during each month. Dark horizontal bars indicate the approximate dates of each sport season, including playoffs and finals.}
    \label{fig:analogy_results}
\end{figure}

\subsection{Subreddit Analogies}
In order to evaluate and tune monthly snapshot community embeddings, we use the set of subreddit analogies provided by Waller and Anderson\cite{Waller2021-yq}, which covers global university-city and city-team pairings for the four major North American sports leagues (MLB, NBA, NFL, NHL). The models are tuned to correctly retrieve appropriate responses, like \textsf{pittsburgh} when presented with  \textsf{Buffalo - buffalobills + steelers = ?}.
Some analogies have multiple correct answers. For instance, New York City has two baseball teams, the Yankees and the Mets. So an analogy is considered correct if the expected subreddit is within the top 5 embeddings nearest the computed result by cosine similarity. We report the proportion of correct analogies out of the total possible solvable analogies for the month, following this Precision@5 definition used by Waller and Anderson\cite{Waller2021-yq}. If any subreddit involved in an analogy is missing from the month's data snapshot, that analogy cannot be solved and is excluded for that month, as reflected in the ratios of correctly solved analogies out of the total possibly solvable in figure \ref{fig:analogy_results}. Good performance on the subreddit analogy task is our first measure of quality and indication that the monthly snapshot embeddings are consistently learning known relationships between subreddits. 

\begin{figure*}[t]
    \begin{tabular}{lll}
        \includegraphics[scale=0.48]{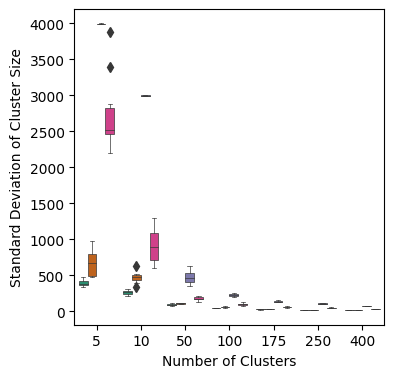} &
        \includegraphics[scale=0.48]{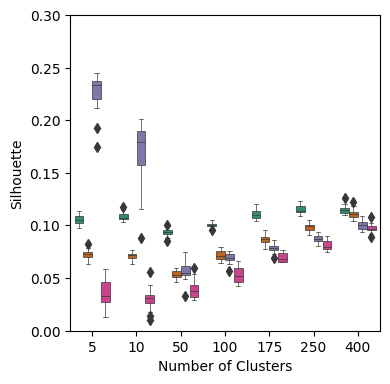} & \includegraphics[scale=0.48]{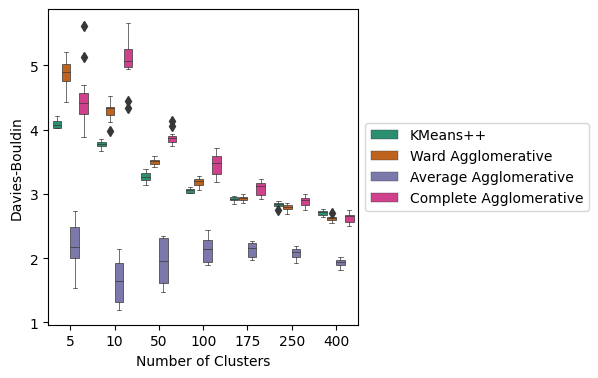} \\
    \end{tabular}
    \captionof{figure}{Intrinsic measures of quality of clustering models, where a single model for each type is trained from each month's snapshot embedding for varying numbers of clusters, showing plots of scores averaged over the year. Higher is better for Silhouette and lower is better for Davies-Bouldin.}
    \label{fig:quantiative_metrics}
\end{figure*}

\section{Methodology}

\subsection{Community Embeddings}
Underlying community embeddings is the idea that a user's comment in one subreddit can be used to predict whether they've commented in other subreddits in the same time period. Training a neural network to do this prediction results in vector weights for each subreddit. This is parallel to word embeddings for natural language, with a user analogous to the context window and the subreddits they comment in analogous to words. 

Our strategy for training community embedding models closely follows Waller and Anderson\cite{Waller2021-yq}, with one key difference: By treating each month of Reddit comments as a snapshot, we train separate embeddings for each month $t$, allowing each monthly model's parameters to be tuned for high performance on solving subreddit analogies, rather than training a single model on the full time range of data. Like Waller and Anderson, we trained each $t$-th skip-gram model on month $t$'s snapshot using negative sampling for efficient approximation, random downsampling of high frequency words\cite{Mikolov2013DistributedRO}, and an "infinite-sized window", where all of each user's comments are used to generate skip-grams. Our models are trained with the Gensim Word2Vec implementation\footnote{\url{https://radimrehurek.com/gensim/models/word2vec.html}} and for each month $t$, we use grid-search to optimize the model parameters as follows:
\begin{itemize}
\item Negative samples $k$: 10, 20
\item Threshold for downsampling high frequency subreddits: 0, 0.001, 0.005
\item Learning rate: 0.05, 0.08
\end{itemize}
Initially, we also experimented with vector dimensions for embeddings, but found that vector size 100 generally performed well. All models are trained for 5 epochs. 
Performance of the best model for each month's snapshot is plotted in figure \ref{fig:snapshotstats} and broken down in detail by type of analogy in figure \ref{fig:analogy_results}. Our monthly snapshot models average 0.64 P@5 compared to 0.94 reported by Waller and Anderson using a model trained on 13 years of Reddit comments\cite{Waller2021-yq}. All subsequent results in this paper are generated from the community embedding that performed best on the analogy task in a given month. For the purposes of this paper, we use L2-normed vectors and cosine distance to create clusterings of subreddits. 

\subsection{Clustering Models}\label{sec:ClusteringModel}
Snapshot embedding models can be used to directly examine nearest neighbors of a subreddit of interest, but unsupervised clustering adds value when the resulting groupings align with end users' goals. In the ideal case, users are presented with coherent clusters of subreddits relevant to their information need for a time period without needing to specify any additional input parameters, then they can see how those clusters change in time. 

We determine how well clusterings meet the information needs of social media researchers by comparing the properties of the following unsupervised clustering models created by taking the snapshot community embedding of month $t$ as input features to produce a clustering $\mathcal{C}_t$:

\textbf{K-Means++} is a widely-used method of clustering which minimizes the average distance of data points to their cluster's center. Greedy k-means++ selects initial centroids probabilistically weighted by data point distributions and includes multiple trials when selecting centers, avoiding the pitfall of selecting "bad" centers that would lead to local minima\cite{Arthur2007kmeansTA}. Choosing different initial centers may result in different clusterings. Granularity is controlled by specifying the desired number of clusters for each use case. 

\textbf{Hierarchical Agglomerative Clustering} (HA) recursively merges clusters closest together to build a hierarchical tree, grouping data points in a bottom-up fashion. Different link criteria can be used to change how clusters are merged at each recursive step. We experimented with \textbf{Ward-linkage}, which minimizes the variance between clusters, \textbf{average linkage}, taking the average distance between data points in clusters, and \textbf{complete linkage}, where distance between clusters is determined by the maximum pairwise distance between data points\cite{Jain1999DataCA}. Granularity is adjusted by setting a desired number of clusters or a maximum distance allowed between merge-able clusters.
Two traits make hierarchical clustering particularly appealing. First, it produces an easy-to-navigate hierarchy of the data. Second, so long as a consistent strategy is used to break ties when distances are equal, it is deterministic, not sensitive to model initialization parameters, and easy to reproduce.
 
\section{Clustering Evaluations} 
Evaluating the quality of clusterings is a notoriously subjective problem, dependent on features of a particular dataset and the context in which the clusters will be used\cite{Von_Luxburg2012-um}. Since our goal is to support a broad range of exploratory and confirmatory analyses of Reddit, we rely on both qualitative and quantitative approaches to describe characteristics of clusterings produced from community embeddings. Intrinsic metrics and statistical measures can be used as a proxy for human judgements, but we prefer coupling them with human judgements to understand which clustering properties users prefer in this setting. 

\subsection{Intrinsic Measures of Cluster Quality}
These measures capture mathematical properties of the groupings of data points to describe \textit{compactness}, how dense data points are in a single cluster, and \textit{separation}, how much distance is between different clusters. 

\textbf{Silhouette Coefficient} is calculated for each data point by taking a ratio of its average distances to data points in the same cluster to the nearest data point in a different cluster. Silhouette Coefficients of data points are averaged to get an overall score for a clustering model. Values range from -1 to 1, where higher values are indicative of better clustering assignments and scores around 0 mean clusters overlap\cite{ROUSSEEUW198753,Tan2022IntroductionTD}. 

\textbf{Davies-Bouldin} is computed by comparing a cluster's separation from its most similar neighboring cluster. Lower scores correspond to better clusterings, indicating within-cluster data points are tight around centroids while the two centroids are separated. The scores for all clusters are averaged as a final score for the clustering. Davies-Bouldin scores have a minimum of 0 and no upper-bound\cite{Davies1979ACS}.  

Although both metrics are designed to measure similar cluster characteristics, there can be different trade-offs depending on the data and use case for clustering. We use the scikit-learn implementations of both measures\footnote{\url{https://scikit-learn.org/stable/modules/classes.html\#clustering-metrics}}.

\begin{figure*}
    \includegraphics[width=0.95\textwidth, height=0.35\textheight]{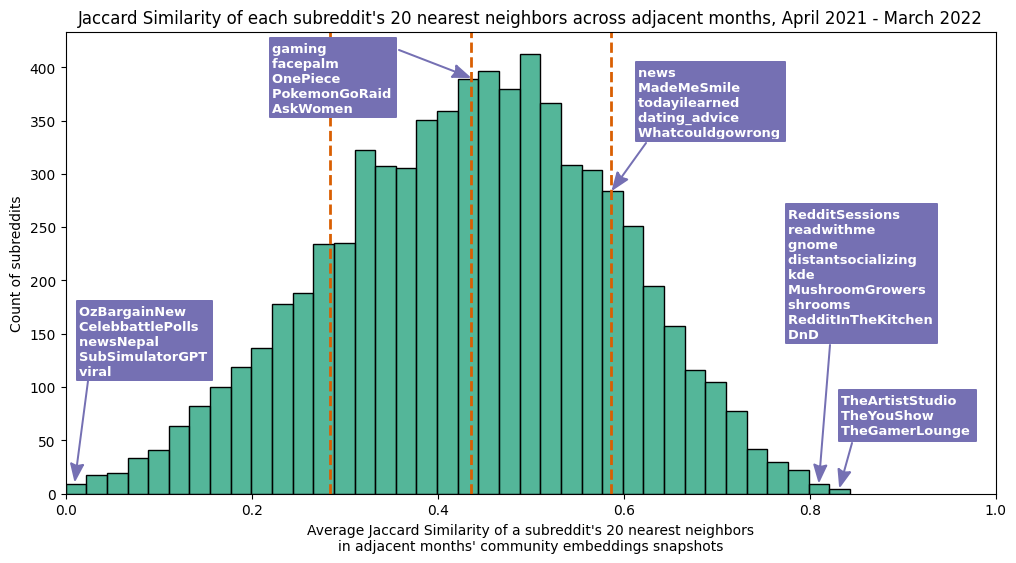}
    \captionof{figure}{Histogram of average Jaccard Similarity of each subreddit's 20 nearest neighbors under community embedding snapshot models in adjacent months. Vertical dashed lines indicate mean and one standard deviation above and below. The most stable communities, including Reddit Public Access Network, appear on the right.} 
    \label{fig:jaccard_sim}
\end{figure*}

\subsection{Qualitative Evaluations} \label{sec:annotation_tasks} By gathering input from human experts, we seek to understand which properties of clusterings users prefer. These experiments can be understood as an initial usability study on the different clustering methods, helping us decide which clustering models to deploy in the live webtool. Two social media researchers and one digital collections librarian, all regular Reddit users, were each presented with clusterings from monthly community embeddings. Would they find the groupings sensible and trustworthy enough to use the models as an exploratory tool? Additionally, instances where annotators disagree with the model or each other provide insight on ways to improve the methodology and areas of inherent ambiguity in the Reddit space.  

\textbf{Cluster coherence} serves as our first qualitative evaluation of cluster quality.  Independently, each annotator marked a cluster as \textit{coherent} if the subreddits assigned to it had an identifiable theme, then named the theme. For example, a cluster consisting of the subreddits \textsf{canada, ontario, PersonalFinanceCanada, vancouver, ottawa} might have a \textit{Canada} theme. This task is also the first step in a grounded theory approach\cite{Blandford2016QualitativeHR} to understanding subreddit communities, helping us to see which themes re-occur in different months. These themes would eventually be used to build the webtool's hierarchical taxonomy of subreddits, but we leave that discussion for future work.

The \textit{coherence score} of a cluster $C_i$ in $\mathcal{C}_t$, $\mathrm{CS}_{C_i}^{\mathcal{C}_t}$, is the percentage of annotators who marked it as coherent. This can be averaged over all clusters to obtain the \textit{clustering coherence score}: $$\mathrm{CS}_{\mathcal{C}_t} = \frac{1}{|\mathcal{C}_t|} \sum_{C_i \in \mathcal{C}_t} \mathrm{CS}_{C_i}^{\mathcal{C}_t}$$
A clustering's overall usefulness can also be upper-bounded by the share of clusters that \textit{at least one annotator} found coherent: $$\textrm{C-Upper}_{\mathcal{C}_t}= \frac{|\{\mathrm{CS}_{C_i}^{\mathcal{C}_t} > 0 \}|}{|\mathcal{C}_t|}$$
For situations where user trust is critical, the share of clusters that \textit{all annotators} found coherent may be more appropriate: $$\textrm{C-Lower}_{\mathcal{C}_t}= \frac{|\{\mathrm{CS}_{C_i}^{\mathcal{C}_t}= 1 \}|}{|\mathcal{C}_t|}$$

If annotators mark many clusters as coherent, we believe end users will find the clustering model's outputs sensible and have trust in the system's exploratory power, even if not every cluster is useful to their work. We measure inter-annotator agreement using Gwet's AC1\cite{Gwet2008ComputingIR} to gauge the level of subjectivity for this task. 

\textbf{Subreddit intruder detection} builds on a word intrusion task originally used to measure the coherence of topic models\cite{Chang2009ReadingTL}, and involves presenting a set of six subreddits to annotators. Five of the subreddits are clustered together by the model, but the sixth is randomly selected, then a random arrangement of the subreddits is presented to annotators. If all annotators are able to pick out the random intruder, the subreddit cluster can be considered coherent. For example, in the set \textsf{PokemonGoFriends, religion, pokemongo, PokemonGoRaids, pokemontrade, pokemon}, the clear non-Pok\'emon intruder is \textsf{religion}.
If annotators do not agree on a single intruder, they may be choosing randomly because the cluster is nonsensical or lacks coherence. Alternatively, annotators might agree on a subreddit to be the intruder which was not actually randomly inserted. In that case, the cluster might be ambiguous or cover multiple, overlapping themes. 

To avoid subreddits' relative popularity misleading annotators during this task, we restricted the selection of the random intruder to subreddits that had a similar number of comments during the time period. We first calculated the standard deviation of the number of comments in the 10K subreddits included in the month $t$ snapshot, $\sigma_t$. For each cluster $C_i$ in a clustering $\mathcal{C}_t$, we selected the top five most popular subreddits in $C_i$ by number of comments that month. Taking $\mu_i$ as the average number of comments for those five subreddits, an intruder $\mathbf{I}_{C_i}$ is drawn from all the subreddits not in $C_i$ with $\mu_i \pm \sigma_t$ number of comments during time $t$. Clusters with fewer than five subreddits or no valid intruder available were not used for this task. 

As in the original word intrusion task\cite{Chang2009ReadingTL}, \textit{model precision} measures how frequently annotators picked out the random intruder for cluster $C_i$ in the clustering $\mathcal{C}_t$. 
The model precision is formulated as $$\mathrm{MP}_{C_i}^{\mathcal{C}_t} = \frac{\textrm{number of annotators to identify } \mathbf{I}_{C_i}}{\textrm{total number of annotators}}$$

\begin{figure*}[t]
    \includegraphics[width=\textwidth]{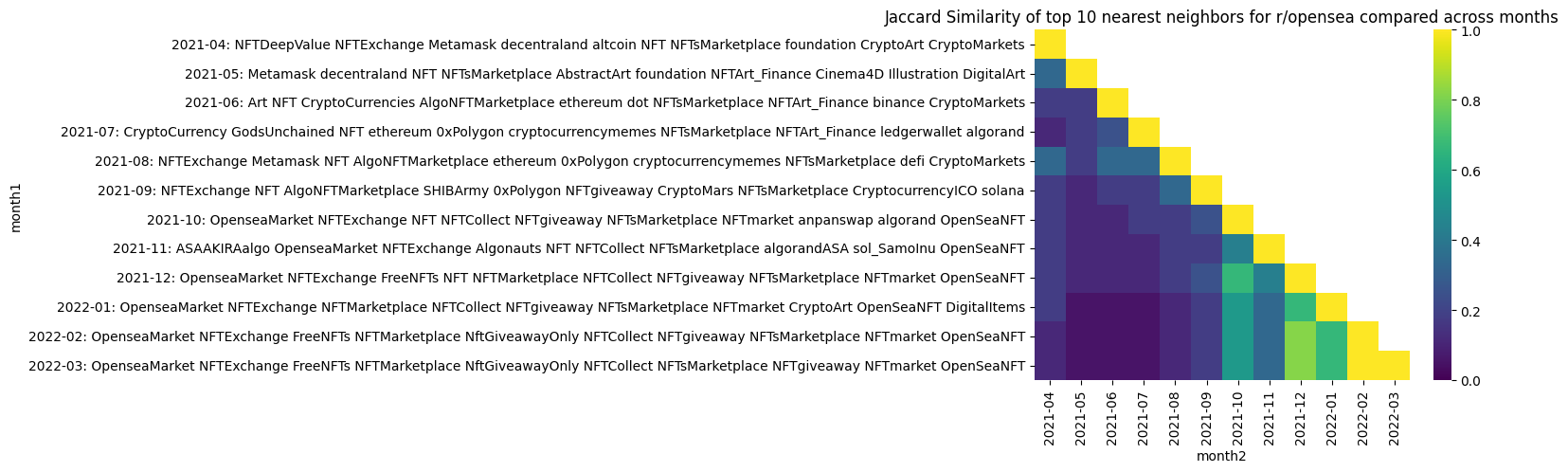} 
    \captionof{figure}{The nearest neighbors of \textsf{r/opensea} become more stable during the data's time frame, starting near a variety of subreddits devoted to art and crypto and ending amongst subreddits devoted to NFT trading.}
    \label{fig:opensea}
\end{figure*} 

\subsection{Temporal Stability} 
\label{section:vi_definition}
Similarity of snapshot models from one month to the next is important for the usability of our models. Consider how topographic maps change gradually because of erosion, but a geopolitical map shows drastic changes based on events like the fall of the Soviet Union. Similarly, we expect most changes on Reddit to be gradual over time, but drastic changes must still be obvious. If month-to-month changes are too drastic, the webtool can't  serve its purpose, because users would be overwhelmed trying to distinguish changes due to noisy data from meaningful shifts in Redditors' behavior and our taxonomy of subreddits would be difficult to maintain. Measures of temporal stability show that monthly snapshots can indeed be used to discover shifts in communities and clusterings change gradually, but the choice of clustering method affects the temporal consistency of groupings.

\textbf{Jaccard Similarity} compares membership of two sets, $A$ and $B$, ranging from 0 when the sets have no elements in common to 1 when they contain exactly the same members. 
$$J(A, B) = \frac{|A \cap B|}{|A \cup B|}$$

Using Jaccard Similarity in pairwise comparisons of a particular subreddit's nearest neighbors across months, we can describe how the community active in that subreddit may be shifting. 

\textbf{Variation of Information} (VI) is an information theoretic metric that can compare clusterings of the same data points, subreddits in our case. In simple terms, VI measures how much information is required to turn one clustering into another. If only a few data points need to be reassigned, the clusterings are similar and VI is low. To compare two clusterings $\mathcal{C}_i$ and $\mathcal{C}_j$, VI is computed using entropy $H$ and mutual information $I$ as 
$$VI(\mathcal{C}_i, \mathcal{C}_j) = H(\mathcal{C}_i) + H(\mathcal{C}_j) - 2I(\mathcal{C}_i, \mathcal{C}_j)$$

Several properties of VI make it appealing for our use case. First, it is a metric on clusterings and follows all the axioms of a distance metric, namely non-negativity, symmetry, evaluating to zero only when clusterings are identical, and obeying the triangle inequality.
Moreover, VI does not directly depend on the number of data points in the data set, so VI on clusterings of differently sized data sets can be interpreted on the same scale, as long as there is a fixed upper bound on the number of clusters\cite{Meil2007ComparingCI}. In our case, $\max(|\mathcal{C}_i|, |\mathcal{C}_j|)=101$ gives an upper bound on $VI(\mathcal{C}_i, \mathcal{C}_j)$ of 13.32. VI allows us to measure the differences between subreddit clusterings in different temporal snapshots, even when different numbers of subreddits overlap between months. 

When we want to compare a clustering $\mathcal{C}_i$ of subreddits $S_i$ to a clustering $\mathcal{C}_j$ of subreddits $S_j$, we extend the clusterings to also account for subreddits which only appear once by assigning subreddits which did not originally appear in time period $t_i$ to a new cluster, $C'_i = \{s | s \in S_j, s \notin S_i\}$, then define $\mathcal{C}'_i = \mathcal{C}_i \cup \{C'_i\}$. Similarly,  $C'_j = \{s | s \in S_i, s \notin S_j\}$ and $\mathcal{C}'_j = \mathcal{C}_j \cup \{C'_j\}$. Now $\mathcal{C}'_i$ and $\mathcal{C}'_j$ partition the same set of data points, $S=S_j \cup S_i$. This approach allows us to compare clusterings of the top 10K subreddits during different temporal snapshots of Reddit, measuring the stability of clusterings as a user would experience them, seeing clusters merge or shift over time as user comment behavior changes or subreddits change in popularity.

\begin{table*}[t]
\centering
\resizebox{.95\textwidth}{!}{
\begin{tabular}{|c|c|c|c|c|c|c|c|} \hline
Month & Clustering Model Type & Silhouette & Davies-Bouldin & Clustering Coherence & C-Upper & C-Lower & Average Model Precision  \\ \hline
\multirow{2}{*}{2021-07} & Average Agglomerative & 0.0689 & \textbf{2.42} &  0.61 & 0.87 & 0.44 & 0.8087  \\
 & K-Means++ & \textbf{0.0989} & 3.11 &  \textbf{0.87} & \textbf{0.96} & \textbf{0.76} & \textbf{0.9023}  \\ \hline 
\multirow{2}{*}{2022-03} & Average Agglomerative & 0.0684 & \textbf{2.26} &  0.72 & 0.89 & 0.48 & 0.8249  \\ 
 & K-Means++ & \textbf{0.1092} & 2.99 &  \textbf{0.90} & \textbf{0.97} & \textbf{0.81} & \textbf{0.8961}  \\ \hline 
\end{tabular}
}
    \caption{Clustering evaluation results for clusterings built from monthly snapshot embeddings for July 2021 and March 2022. Annotators strongly preferred the k-means++ models across all qualitative evaluation tasks. Bold indicates better values. }
    \label{table:annotation_stats}
\end{table*}

\section{Experimental Results and Discussion}
\subsection{Temporal Stability}
Even before clustering, the community embedding snapshots reveal temporal trends in analogy performance. As seen in figure \ref{fig:snapshotstats}, community embeddings produced from each month are able to consistently achieve high P@5 on solving subreddit analogies, yet temporal shifts are obvious by category. Collections of subreddits will have more activity during certain times of the year, determining which analogies can be solved, an effect of our data pre-processing approach. Figure \ref{fig:analogy_results} shows that there are more solvable analogies involving teams during each sport's season, which is particularly pronounced for baseball. University-city pairs have more solvable analogies during August, September, and January, times when students may be actively commenting to find housing and information about their school or neighborhood at the beginning of the semester. Even in time periods when there are fewer subreddits from a category retained, the community embedding still solves a high proportion of the remaining analogies.

Changes in an individual subreddit's nearest neighbors from one month to the next offer another way to analyze temporal shifts in communities on Reddit. Figure \ref{fig:opensea} shows the 10 nearest neighbors of \textsf{opensea} each month, a subreddit dedicated to a marketplace for selling non-fungible tokens (NFTs) associated in particular with digital artwork. In the earliest months of our dataset, April through June 2021, \textsf{opensea} had a shifting variety of subreddits in its top 10 nearest neighbors: other NFT marketplaces (\textsf{NFTsMarketplace}), general interest cryptocurrency and blockchain subreddits (\textsf{ethereum, altcoin}), and those dedicated to visual art (\textsf{DigitalArt, AbstractArt, Illustration}). The Jaccard Similarity of \textsf{opensea}'s top 10 nearest neighbors was correspondingly low, since less than a third of the neighbors were shared between months. Over the course of 2021, as popularity and sales of NFTs exploded\cite{NFTsFinTimes}, the art and cryptocurrency focused subreddits disappeared from \textsf{opensea}'s neighbors, replaced by similar sets of subreddits dedicated solely to exchanging NFTs (\textsf{OpenseaMarket, NftGiveawayOnly, NFTCollect, NFTExchange}) from December 2021 on. Our method corroborates the establishment of a Reddit community around exchanging NFTs coinciding with growth in NFT sales. 

To see how this observation would generalize, for the 6,950 subreddits that appeared in every monthly snapshot, we analyzed the Jaccard Similarity of their 20 nearest neighbors in adjacent months. For example, the 20 most similar subreddits to \textsf{aww}, the subreddit dedicated to cute animal pictures, under the April 2021 community embedding model would be compared to \textsf{aww}'s 20 most similar subreddits in May 2021 using Jaccard Similarity, then May 2021 compared to June 2021, and so on. This results in 11 scores for pairs of adjacent months for every subreddit. A histogram of this average Jaccard Similarity is presented in figure \ref{fig:jaccard_sim}. 

At its extreme low value, average Jaccard Similarity identifies subreddits with comments almost exclusively from bots dedicated to that subreddit, making it difficult to anchor them to other subreddits in the embeddings. These include subreddits intended to function like RSS feeds (\textsf{newsNepal, OzBargainBin}), facilitate forum polls (\textsf{CelebbattlePolls}) or observe a chat bot interacting with itself (\textsf{SubSimulartorGPT}). They constitute an interesting category of subreddits which exist only to be observed by Redditors. Because there are too few non-bot user comments, the models cannot learn a position for these subreddits relative to others and give them a random embedding and set of neighbors each month. These subreddits are easy to detect in the temporal models, but one potential improvement is to remove subreddits that don't reach some threshold of unique commenters during pre-processing.

Inversely, high similarity of nearest neighbors over time can be used to identify subreddits that have strong connections within a particular Reddit community during the entire year. The top 5 most stable subreddits found with this method, \textsf{readwithme, RedditSessions, TheArtistStudio, TheYouShow, TheGamerLounge}, all belonged to the Reddit Public Access Network (RPAN), a set of 17 subreddits to which users could live-stream, which launched in 2019\cite{RPANVerge}. In a note to the community in November, 2022, Reddit announced discontinuation of the service citing costs, thanking the RPAN's "diehard fans and avid moderation teams"\footnote{\url{https://www.reddit.com/r/pan/comments/yl5zzd/update_on_the_future_of_live_video_broadcasting/}}. Although we weren't previously aware of RPAN, our method was able to detect this community, which had an active, dedicated user-base in 2021, as people facing lockdowns during the COVID-19 pandemic turned to online socializing. Other subreddits with highly stable neighbors also belonged to easily identifiable communities, namely open source software (\textsf{gnome, kde}) and mushroom cultivation for eating and psychedelics (\textsf{MushroomGrowers, shrooms}).

At less extreme values, this average Jaccard Similarity is difficult to interpret. Popular general interest subreddits, like \textsf{gaming, facepalm} and \textsf{AskWomen}, fall near the average, but so do subreddits which one might expect to be part of a dedicated community, like one for fans of the anime \textit{One Piece} and another for organizing players of the mobile game \textit{Pok\'emon Go}. The value is not strongly correlated with a subreddit's popularity by number of comments in the time period (Pearson's $r=0.063$), but we encourage future work to explore if it is related to other features, such as subreddit's user retention or generalist-specialist score\cite{Waller2019GeneralistsAS}.

\begin{figure}
    \includegraphics[scale=0.42]{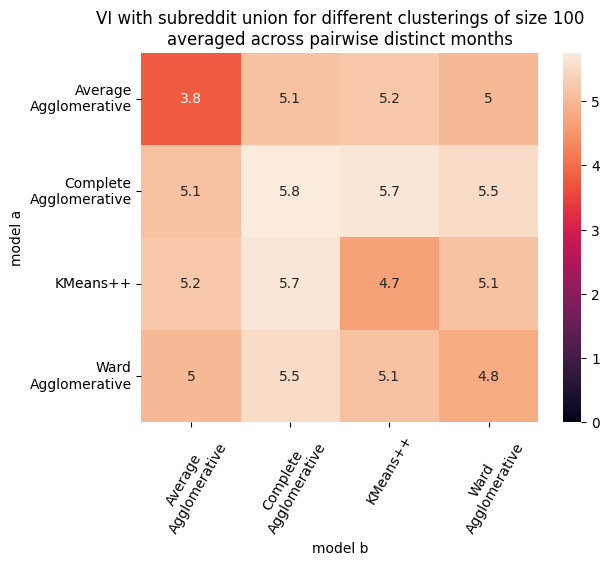}
    \captionof{figure}{ VI to compare clusterings of size 100 across distinct monthly snapshots shows HA average linkage models produce the most stable clusterings across time. Each cell averages $\frac{m(m-1)}{2} = 64$ VI comparisons for $m=12$ months.}
    \label{fig:inter_treatment_vi}
\end{figure}

\begin{figure}
    \includegraphics[scale=0.42]{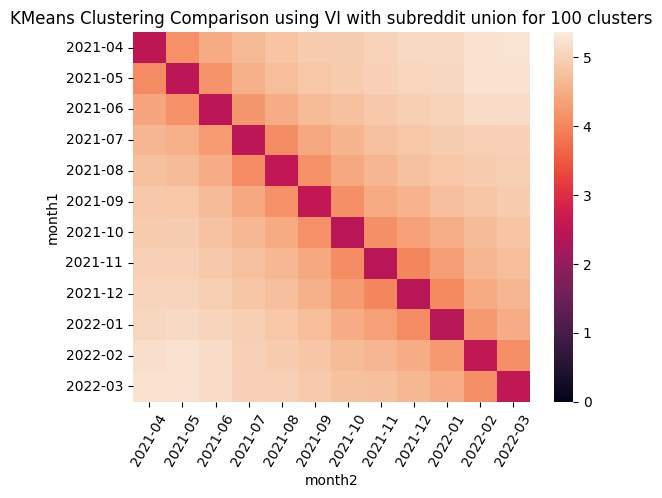}
    \captionof{figure}{VI pairwise comparing 10 k-means++ models for each month with different initialization parameters. Each cell gives the average VI across 45 clustering comparisons between the given months. Changes in clusterings over time are gradual.}
    \label{fig:vi}
\end{figure}

\subsection{Quantitative Properties of Clusterings}
\label{sec:QuantitativeProperties}
Intrinsic and statistical measures describe the properties of clusterings produced by different approaches. By better understanding the clustering behaviour, we can make informed decisions about which "lines to draw on the Reddit map" that will be presented to users of our webtool.  

One major trade-off for usability and clustering quality rests in the cluster size. Clusters containing too many subreddits to identify a useful common theme may not meet users' information needs for general exploration and discovery. Cluster size is affected by both the type of clustering model and the specified level of granularity, the number of clusters. From figure \ref{fig:quantiative_metrics} we can see that k-means++ and HA average linkage models get the best Silhouette and Davies-Bouldin scores across a range of different clustering sizes. HA average linkage appears to work particularly well when splitting the subreddits into only 5 or 10 clusters, although that's fairly coarse-grained for division into topical themes. However, HA agglomerative and complete linkage models also have a high standard deviation in number of subreddits assigned to each cluster.  When the number of clusters was set to 50 or greater, the HA average linkage always produced a cluster containing only 1 subreddit, and largest cluster over all months contained more than double the number of subreddits compared to Ward or k-means++. The HA average and complete linkage models assign almost all the subreddits to one large cluster and shave off outliers into many clusters, which has questionable utility to users, as we'll see in qualitative evaluations.

Applying the elbow heuristic method to determine the appropriate number of clusters\cite{ROUSSEEUW198753} from figure \ref{fig:quantiative_metrics} also suggests that Silhouette and Davies-Bouldin scores do not improve significantly when the number of clusters is greater than 100. We believe that information needs for general exploration and discovery are best met when the number of clusters is around 100, seeking a balance between a tractable number of clusters to browse and the number of subreddits which can be judged at a glance for each cluster. We acknowledge that this is subjective and dependent on each user's goal and familiarity with Reddit data. We leave further discussion of improvements to presentation of clustered  Reddit data to future work. 

\subsection{Stability of Clustering Models} 
We also want to establish how different clustering approaches behave with respect to temporal stability. Using the best community embedding snapshot each month, we trained clustering models with size 100, as described in section \ref{sec:ClusteringModel}, then use VI to pairwise compare clusterings from different months. Summarized in figure \ref{fig:inter_treatment_vi}, this shows the effect of clustering model type on  similarity between subreddit clusterings over time. Clusterings produced within the same type of model are generally the most stable across the entire year, but clusterings produced from different models are comparable and the choice of clustering model and parameters does matter. HA average linkage clusterings are the most similar month-over-month, while HA with complete linkage are the least stable. Clusterings are also fairly stable within k-means++ and HA Ward models. 

We repeated this experiment using only k-means++ models, training 10 models on each monthly snapshot and varying the models' initial starting parameters, to see how sensitive clusterings would be to randomness inherent to k-means++. Low average VI within the same month suggests consistent clusterings can be produced from the community embeddings and the process isn't sensitive to starting parameters. VI of these inter-monthly experiments appear in figure \ref{fig:vi}, where the gradual change of clusterings is clear. The similarity of clusterings from the same month is seen in the dark diagonal band, which fades out to the most different clusterings between months furthest apart in time. To some extent, this relationship is due to our strategy of computing VI based on the union of the top 10K subreddits between months, which also change gradually over time, but this captures the way end-users of the clusterings experience the changes when browsing subreddit clusters. 

\subsection{Qualitative Clustering Evaluations}
\begin{figure}
    \includegraphics[scale=0.4]{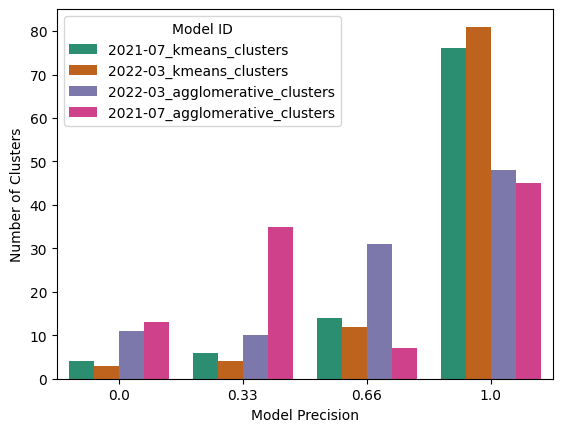}
    \captionof{figure}{Distribution of $\mathrm{MP}_{C_i}^{\mathcal{C}_t}$ across clusters annotated in our intruder task. Annotators were more able to identify the random intruder in clusters from k-means++ models.}
    \label{fig:model_precision}
\end{figure}

Based on their strong intrinsic measures and temporal stability, we focused manual annotation efforts on the HA average linkage and k-means++ clusterings. We arbitrarily selected two monthly snapshots, July 2021 and March 2022, for annotation at a granularity of 100 clusters. In terms of the tasks described in section \ref{sec:annotation_tasks}, annotators marked k-means++ clusters as more coherent and also more readily identified intruder subreddits for k-means++, as summarized in table \ref{table:annotation_stats} and figure \ref{fig:model_precision}. 

Inter-annotator agreement on the coherence task was also higher for k-means++, suggesting that judging coherence for the HA average linkage was more subjective or challenging for annotators. For k-means++ clusterings, Gwet's AC1 was 0.85, falling in the excellent range, while on HA average linkage, it was in the moderate range at 0.50\cite{Wongpakaran2013ACO}. The major source of disagreement were clusterings that HA average linkage assigned only one or two subreddits, related to the large standard deviation in cluster size discussed in \ref{sec:QuantitativeProperties}. One annotator felt that a clustering consisting of only one subreddit was inherently coherent, while others thought this fails to contextualize that subreddit within a thematic community on Reddit.


Anecdotally, annotators reported high trust in the k-means++ clusters, citing groupings of subreddits that reflected linguistic or geographic features. In both months the k-means++ models surfaced both \textit{Canada} (\textsf{canada, ontario, vancouver}...) and \textit{India} (\textsf{Cricket, indiasocial, india, unitedstatesofindia}...) clusters. In March 2022, they noted two separate clusterings for US cities on the East (\textsf{boston, nyc, philadelphia, AskNYC, washingtondc}...) and West coasts (\textsf{LosAngeles, Seattle, Portland, bayarea, sandiego}...), and a German language cluster (\textsf{de, ich\_iel, FragReddit, Austria, Finanzen, germany, mauerstrassenwetten}...). Results for HA average linkage clusterings with size 100 were underwhelming, but if hierarchical clusters are desired, VI results from figure \ref{fig:inter_treatment_vi} suggest that Ward-linkage produces clusterings similar to k-means++.

\section{Concluding Discussion}
We presented a method for building monthly snapshot community embeddings for popular subreddits that facilitates exploration of Reddit during that time period through similarity relationships and clustering. Despite each month's snapshot being created independently from other months in the dataset, these models capture gradual changes in Reddit over time and can be used in confirmatory and exploratory analyses to understand community trends on Reddit. Cluster coherence and subreddit intruder annotation tasks showed clusterings align with the judgements of human experts, especially when using k-means++ models. Our monthly snapshot models are available to use via a web interface at \url{RedditMap.social}, which we hope will provide an intuitive way to browse and better understand discourse on Reddit. We look forward to seeing how fellow researchers use the tool and heartily welcome suggestions for improvement. 

Although recently introduced access restrictions on the Reddit API\cite{APIVerge}, including policy changes around the Pushshift dataset, are hindering updates to the webtool, we are closely following developments and plan to bring the models up to date as soon as possible. Like most researchers focused on Reddit, we see Pushshift as an invaluable research tool, and we may be forced to significantly change our methods if Pushshift and Reddit are not able to find agreement on methods for archiving and studying the site. We are especially interested in studying the impact of the 2023 protests that swept across Reddit in response to the company's monetization of its API. 

We also anticipate many ways to improve the tool and models, such as applying soft clustering to create multiple lenses to view relationships between subreddits and incorporating genealogy graphs or procrustes alignment to further compare monthly snapshot embeddings. Additional features, especially text, Reddit karma or user growth or retention, may also be integrated to generate further insights.

\bibliographystyle{ACM-Reference-Format}
\bibliography{ihop_reddit} 

\end{document}